\documentstyle[12pt]{article}

\begin{document}

\setcounter{page}{1}

\centerline{\large {\bf The Effective Action for QCD
}}
\centerline{\large {\bf at High Energies\footnote{Talk given
at the "Theory of Hadrons and Light-front QCD" workshop in
Zgorzelisko, Poland, 15-25 Aug. 1994.}
}}
\vskip.3in
\centerline{ Lech Szymanowski\footnote{E-mail: lechszym@fuw.edu.pl} }
\centerline{ Soltan Institute for Nuclear Studies }
\centerline{ Hoza 69, 00-681 Warsaw, Poland }

\vskip.6 in
\centerline{\bf Abstract}
\vskip.1in
I discuss the construction of the effective action for QCD suitable
for the description of  high-energy and small momentum transfer
diffractive processes.
\vskip.1in

\vskip.3in
{\bf 1. Introduction}
\vskip.1in
In my talk I discuss the results obtained in collaboration with Roland
Kirschner and Lev Lipatov  which are the subject of Ref.~1
 and Ref.~2.

As we learned from the talk by J. Bartels$^3$ in new
experiments at HERA on electron-proton deep inelastic scattering
one can probe the region of very small values of the Bjorken variable
$x = -\frac{Q^2}{s} \sim 10^{-4}$, where $s$ is the scattering
energy squared and $Q^\mu$ being the momentum transferred by the
photon. The results of the experiments show that the gluonic structure
functions increase fairly strong for small $x$. Such a behaviour seems
to be in agreement with the theoretical predictions based on the
BFKL equation$^4$. This equation is obtained in the leading
logarithm approximation (LLA) which corresponds to the sum of
perturbative contributions being series in the effective coupling
constant $g^2 \ln\frac{1}{x} \ll 1$( $g$ is the QCD coupling constant).

\noindent It is known that LLA violates unitarity so the growth of
the number of gluons in the nucleon cannot continue forever. This
means that one should apply the unitarization procedure which restores
the unitarity by taking into account the screening effects and which
goes beyond LLA.

 The  method of unitarization proposed by L. Lipatov$^5$
 is based on the use of the effective lagrangian for QCD
at high-energies and small momenta transfer i.e. in the multi-Regge
kinematics (MRK). This lagrangian being simpler than original QCD
lagrangian contains all important physical modes which are present
in MRK. Its form reflects also the relationship of the four-dimensional
QCD at high-energies with two-dimensional theories related to
two-dimensional space of transverse momenta and the two-dimensional
space of longitudinal momenta.

One can derive the effective lagrangian for QCD in several ways.
Using the diagrammatic method (see Ref.~2) we derive from tree
graphs in MRK the effective vertices for scattering and production
of gluons and quarks. Then by an appropriate choice of the quark wave
function and the gluon polarization vectors  these vertices can
be represented in a simple form.

\noindent One can also start directly from the original QCD
lagrangian and try to eliminate  modes of gluons and quarks which are
not present in MRK. This elimination procedure can be performed
either in the framework of the path integral formalism (see Ref.~2) or
by means of the equations of motion (see Ref.~1). Below I shall
describe the last method and for the simplicity of presentation we
 consider only the gluonic part of the QCD lagrangian.

\vskip.3in
{\bf 2. Sketch of construction of the effective lagrangian}
\vskip.1in

Because of the MRK it is natural to work in the light-cone gauge
defined by one of the momenta $p_A$ or $p_B$ of the initial
massless scattering particles. In the c.m.s. where
\begin{equation}
\label{1}
p^0_A = p^0_B = p^3_A = -p^3_B =
\frac{\sqrt{s}}{2} ,\qquad p_{A\perp} = p_{B\perp} = 0
\end{equation}
we choose for the definiteness the light-cone gauge
\begin{equation}
\label{2}
   A_- \equiv A_0 - A_3 = p^{\mu}_B A_{\mu} = 0 \;\;\;.
\end{equation}
The gluonic part of the QCD lagrangian
\begin{equation}
{\cal L} = - \frac{1}{4}G^a_{\mu\nu}G^{a\mu\nu} \;\;\;,\;\;\;
G^a_{\mu\nu} = \partial_{\mu}A^a_{\nu} - \partial_{\nu}A^a_{\mu}
     + gf^{abc}A^b_{\mu}A^c_{\nu}
\label{3}
\end{equation}
depends quadratically on $A^a_+ = A^a_0 + A^a_3$ so one can
eliminate this variable by means of the equations of motion.

\noindent In such a way we arrive to the lagrangian which
depends only on transverse components $A^a_{\varrho}, \varrho = 1,2$
of the  four-potential
\begin{eqnarray}
\label{4}
{\cal L} &=& \frac{1}{2} A^a_\sigma \Box A^{a \sigma} -
ig (\partial_-
  A^\sigma) T^a A_\sigma (\frac{1}{\partial_-}
  \partial_\varrho A^{a\varrho}) \nonumber \\
  && - \frac{g^2}{2} (\partial_- A_\sigma) T^a A^\sigma
 \frac{1}{\partial^2_-}
 (\partial_- A_\varrho) T^a A^\varrho - ig (\partial_\varrho A^a_\sigma)
  A^\varrho T^a A^\sigma \nonumber \\
&& + \frac{g^2}{4} A_\varrho T^a A_\sigma A^\varrho T^a A^\sigma
\end{eqnarray}
where $x^{\mp} = x^0 \mp x^3 \;,\; \partial_{\mp} =
\frac{\partial}{\partial x^{\mp}} \;,\; \Box =
4\partial_-\partial_+ + \partial_\varrho\partial^{\varrho}$ and
$T^a$ denotes the generators of gauge group in the adjoint
representation.

\noindent In Eq.~(\ref{4}) as it stays the fields contain
all modes. On
the other hand in the MRK the strongly virtual (heavy)
modes of the fields $A^{(s)}_{\varrho}$
\begin{equation}
\label{5}
k^2
   \simeq k^2_\|
       \gg |k^2_{\perp}|
\end{equation}
are not present since they are already integrated out and we
left with the moderately virtual fields $A^{(m)}_{\varrho}$. This
elimination of the heavy modes is performed within perturbation
theory and in the following analysis we shall restrict ourselves
to the first perturbative order.

Let us decompose $A_{\varrho}$ as the sum of strongly and
moderately virtual fields
\begin{equation}
\label{6}
   A_{\varrho} = A^{(s)}_{\varrho} + A^{(m)}_{\varrho} \;\;\;.
\end{equation}
Substituting the decomposition (\ref{6}) to the lagrangian
(\ref{4}) and neglecting the interference contribution between
$s$- and $m$- fields we obtain as a kinetic part
\begin{equation}
\label{7}
 {\cal L}^{ kin} \cong  2 A^{(s)}_\sigma \partial_+
      \partial_- A^{(s)\sigma} + \frac{1}{2} A^{(m)}_\sigma \Box
         A^{(m) \sigma} \;\;\;.
\end{equation}
As an interaction lagrangian for $s$-fields ${\cal L}^{(s)}$ we
take those terms from the lagrangian (\ref{4}) (after inserting
(\ref{6})) which contain the enhancement factor in the MRK being
the operator $\frac{1}{\partial_-}$ acting on the field with the
smallest $k_-$ momentum component. The resulting lagrangian has
the form
\begin{eqnarray}
\label{8}
   {\cal L}^{(s)} &=& 2 A^{(s)a}_\varrho \partial_+ \partial_- A^{(s) a
 \varrho} \nonumber \\
  && +ig [ A_\sigma^{(m)} T^a \partial_- A^{(s) \sigma} +
  A_\sigma^{(s)}T^a\partial_-A^{(m) \sigma}](\frac{1}{\partial_-}
 \partial_\varrho A^{(m)a
     \varrho}) \;\;\;.
\end{eqnarray}
The  integration over $A^{(s)}_{\varrho}$ fields in Eq.(\ref{8}) leads
to the expression
\begin{equation}
\label{9}
  \Delta {\cal L}
 = \frac{g^2}{4} A^{(m)}_\varrho T^a (\partial_- A^{(m)\varrho})
  (\frac{1}{\partial_+\partial_-}
  \partial_\sigma A^{(m)\sigma} ) T^a
   (\frac{1}{\partial_-} \partial_\eta A^{(m)\eta}) \;\;\;.
\end{equation}
The sum of lagrangian (\ref{4}) involving only $m$-fields and formula
(\ref{9}) leads to the modified lagrangian ${\cal L}^{mod}$
\begin{equation}
\label{10}
   {\cal L}^{mod} = {\cal L}
  |_{A\rightarrow A^{(m)}}
 + \Delta {\cal L} \;\;\;.
\end{equation}
We should never forget about the underlying MRK in which the
$k_-$ momentum components of the  particles are ordered.
In the case of Eq.(\ref{9}),
 the $k_-$ momentum
components of the first two fields $A^{(m)}_\varrho$ are much bigger
than the corresponding ones of the two last $A^{(m)}$'s .

After removing the heavy modes we separate the modes of
$A^{(m)}_\varrho$ into a part involving Coulombic modes
$A'_\varrho$ obeying
$\mid k_+k_-\mid \ll \mid k_{\perp}^2\mid$ and the part
describing the produced particles $A_\varrho$ with
momenta satisfying $\mid k_+k_-\mid \approx \mid
k_{\perp}^2\mid$ (for which we keep the original notation). The
kinetic term of the Coulombic modes involves only transverse
derivatives so these modes describe the instantaneus Coulomb
interaction.

With the help of the Coulombic modes $A'_\varrho$ we can rewrite $\Delta
{\cal L}$ from Eq.(\ref{9}) as a product of the triple vertex
from ${\cal L}$ (Eq.(\ref{4}))
\begin{equation}
\label{11}
 -i g (\partial_-A^{(m)}_\sigma)T^a A^{(m) \sigma}(\frac{1}{\partial_-}
 \partial_\varrho A'^{a \varrho})
\end{equation}
and the induced vertex $\Delta {\cal L}^{ind}$
\begin{equation}
\label{12}
 \Delta {\cal L}^{ind} = \frac{ig}{4} (\partial_- \partial_\varrho
   A'^{a\varrho}) \left( \frac{1}{\partial_+ \partial_-}
  \partial_\sigma A^{(m)\sigma} \right) T^a \left( \frac{1}{\partial_-}
    \partial_\eta A^{(m)\eta} \right)
\end{equation}
connected by the Coulombic propagators resulting from the
kinetic part ${\cal L}_{Coul}^{kin}$
\begin{equation}
\label{13}
{\cal L}_{Coul}^{kin} = \frac{1}{2} A_\varrho^{' a}\partial_\sigma
\partial^\sigma A^{' a \varrho} \;\;\;.
\end{equation}
Let us also note that the remaining terms resulting from the
integration over $A'_\varrho$ fields in Eqs.~(\ref{11}),(\ref{12})
and (\ref{13}) cancel the third term in ${\cal L}(A^{(m)})$
given by Eq.~(\ref{4}).

\noindent If we neglect the last nonsingular term in
Eq.(\ref{4}) we can write the effective lagrangian in the form
\begin{eqnarray}
\label{14}
 {\cal L}^{eff} &=& \frac{1}{2} A^{(m)a}_\varrho \Box
 A^{(m)a\varrho}
  -ig (\partial_-  A^{(m)}_\sigma) T^a  A^{(m)\sigma}
  (\frac{1}{\partial_-}\partial_\varrho A^{'a\varrho})
   \\
  && - ig  (\partial_\varrho A^{(m)a}_\sigma) A^{(m)\varrho}
     T^a A^{(m)\sigma}
    + \frac{ig}{4} (\partial_- \partial_\sigma A'^{a\sigma})
   (\frac{1}{\partial_+ \partial_-} \partial_\varrho A^{(m)\varrho})
     T^a (\frac{1}{\partial_-} \partial_\eta A^{(m)\eta}) \;\;\;.
     \nonumber
\end{eqnarray}

It is convenient now to introduce the following notation for the
Coulombic fields
\begin{equation}
\label{15}
A_+ = - \frac{1}{\partial_-} \partial_\sigma A'^\sigma \;\;\;\;
  A_- = -2 \frac{\partial_-
  \partial_\sigma}{\partial_\varrho
  \partial^\varrho} A'^\sigma \;\;\;.
\end{equation}
According to the above definitions the fields $A_{\pm}$
are dependent. Nevertheless  we declare them in the following as being
independent ones. This is needed in order to put effectively to
zero the term arising from the integration over $A'_\varrho$
discussed above and which cancels the corresponding term of the
order $g^2$ from Eq.(\ref{4}). In such a way we arrive to the
following kinetic part of the gluonic effective lagrangian
\begin{equation}
\label{16}
 {\cal L}_{kin} = \frac{1}{2}A^a_+\partial_\sigma\partial^\sigma
 A^a_- + \frac{1}{2} A^a_\sigma \Box A^{a\sigma} \;\;\;.
\end{equation}
Let us emphasize that the coefficient in the front of Coulombic
part differs from the one obtained by the substitution of Eq.~(\ref{15})
to Eq.~(\ref{13}). It is fixed by requirement that the
amplitudes obtained with the help of $A'_\varrho$ fields coincide with
the amplitudes calculated with the use of $A_{\pm}$ fields.

{}From the effective lagrangian (\ref{14}) one can also read off the
interaction terms. We substitute in Eq.~(\ref{14}) the decomposition
$A^{(m)}_\rho = A'_\rho + A_\rho$ supplemented by introduction
of the definitions (\ref{15}). It is convenient to represent the
result in the complex coordinates and using the analogous notation
for the produced fields
\begin{equation}
\label{17}
 \begin{array}{llll}
 \varrho = x^1 + ix^2, & \varrho^* = x^1 - ix^2, &
    \partial = \frac{\partial}{\partial\varrho} , & \partial^*
    = \frac{\partial}{\partial \varrho^*}  \\
   A = A^1 + iA^2 & A^* = A^1 - iA^2 &  . &
          \end{array}
\end {equation}
Moreover, we describe the produced particles in terms of the complex
scalar fields $\phi^a$ (see Ref.5)
\begin{equation}
\label{18}
   A = i \partial^* \phi\;\;,\;\; A^* = - i \partial \phi^* \;\;\;.
\end{equation}
After that the gluonic effective lagrangian is given as the sum
\begin{equation}
\label{19}
{\cal L}^{eff} = {\cal L}_{kin} + {\cal L}^{(R)}_{scat} +
{\cal L}^{(L)}_{scat} + {\cal L}_{prod} + {\cal L}_{Coul}  \;\;\;.
\end{equation}
In the sum (\ref{19}) the term ${\cal L}_{kin}$ is obtained from
 Eq.~(\ref{16})
 \begin{equation}
 \label{20}
 {\cal L}_{kin}  =  - 2A^a_+\partial\partial^* A^a_-
 - \frac{1}{2}(\partial^*\phi^a)
  \Box (\partial\phi^{a *}) \;\;\;.
 \end{equation}
 The term ${\cal L}^{(R)}_{scat}$ describes the scattering off right
 particles i.e. on the $A_+$ field
 \begin{equation}
 \label{21}
    {\cal L}^{(R)}_{scat} = - \frac{ig}{2}A^a_+        \left[
  (\partial_- \partial^*\phi) T^a(\partial \phi^*) + (\partial_-
 \partial \phi^*) T^a(\partial^* \phi)\right] \;\;\;.
 \end{equation}
 The analogous expression corresponding to the scattering on the
 $A_-$ field ${\cal L}^{(L)}_{scat}$ reads
 \begin{equation}
 \label{22}
    {\cal L}^{(L)}_{scat} = - \frac{ig}{2}A^a_-[
 (\partial_+ \partial^*\phi^*) T^a(\partial \phi) + (\partial_+
 \partial \phi) T^a(\partial^* \phi^*)] \;\;\;.
 \end{equation}
 The term ${\cal L}_{prod}$ describes the production of $\phi$ and
 $\phi^*$
 \begin{equation}
 \label{23}
 {\cal L}_{prod} = g[ \phi^a (\partial A_-)T^a(\partial^*
   A_+) - \phi^{a *}(\partial^* A_-)T^a (\partial A_+) ] \;\;\;.
 \end{equation}
 Finally, the part ${\cal L}_{Coul}$ contains the interaction vertices
 involving the Coulombic fields
 \begin{equation}
 \label{24}
 {\cal L}_{Coul} = \frac{ig}{2}[ (\partial\partial^*
   A^a_-)(\frac{1}{\partial_+}A_+ ) T^a A_+
  +  (\partial\partial^*
  A^a_+)(\frac{1}{\partial_-}A_- ) T^a A_-] \;\;\;.
 \end{equation}
 Summarizing, we constructed the effective lagrangian for QCD from
 which one can reproduce in a very economic way the known results
 about the asymptotics of scattering amplitudes in the MRK and in the
 LLA. The lagrangian (\ref{19}) posses many remarkable properties.
 In particular,
  if we approximate the operator $\Box$ in (\ref{20})
 by $4\partial_+\partial_-$ then the scattering amplitudes
 resulting from the lagrangian (\ref{19}) are given as the
 product of two scattering amplitudes related to the two-dimensional
 theories acting in the longitudinal space and in the transverse
 space.

  \noindent We expect that the effective lagrangian (\ref{19}) is a
 convenient starting point towards construction of a method which
 goes beyond the LLA.

\vskip.3in
{\bf Acknowledgements}
\vskip.1in

This work is partially supported by  Komitet Badan Naukowych
under grant No. 2P302 143 06 and by German-Polish Scientific and
Technological Agreement.

\vskip.3in
{\bf References}
\vskip.1in

\noindent 1. R.Kirschner, L.Lipatov and L.Szymanowski, Nucl.Phys.
{\bf B245},579(1994)   \\
2. R.Kirschner, L.Lipatov and L.Szymanowski, "Symmetry properties
of the effective   \\
\hspace*{0.1in} action for high-energy QCD", DESY preprint 94-064  \\
3. J. Bartels, see these Proceedings   \\
4. L.N. Lipatov, Sov.J.Nucl.Phys. {\bf 23},642(1976) \\
\hspace*{0.1in} V.S.Fadin, E.A. Kuraev and L.N. Lipatov, ZhETF {\bf
 71},840(1976); {\it ibid} {\bf 72},377(1977)  \\
 \hspace*{0.1in} Y.Y. Balitsky and L.N. Lipatov, Sov.J.Nucl.Phys.
 {\bf 28},882(1978) \\
 5. L.N. Lipatov, Nucl.Phys. {\bf B365},614(1991)

\end{document}